\documentclass[useAMS,usenatbib,twocolumn]{mn2e}
\usepackage{times}

\usepackage{graphicx,color}

\input epsf

\usepackage{wrapfig}

\newcommand{\beq}{\begin{equation}}
\newcommand{\beqa}{\begin{eqnarray}}
\newcommand{\eeq}{\end{equation}}
\newcommand{\eeqa}{\end{eqnarray}}

\newcommand{\lsim}{\la}
\newcommand{\gsim}{\ga}
\newcommand{\psim}{\mbox{\raisebox{-1.0ex}{$~\stackrel{\textstyle \propto}
{\textstyle \sim}~$ }}}

\newcommand{\lmk}{\left(}
\newcommand{\rmk}{\right)}

\title{Prospects of eLISA for Detecting Galactic Binary Black Holes Similar to GW150914  } 
\author[]{Naoki Seto
\\
Department of Physics, Kyoto University, 
Kyoto 606-8502, Japan
}

\begin{document}

\maketitle

\begin{abstract}
We discuss the prospects of eLISA for detecting gravitational waves (GWs) from Galactic binary black holes (BBHs) similar to  GW150914. For a comoving merger rate that is consistent with current observation, eLISA is likely to identify at least one BBH with a sufficient signal-to-noise ratio. In addition, eLISA has a potential to measure the eccentricity of the BBH as small as $e\sim 0.02$, corresponding to the residual value $e\sim 10^{-6}$ at 10Hz.  Therefore, eLISA could provide us with a crucial information to understand the formation processes of relatively massive BBHs like GW150914. We also derive a simple scaling relation for the expected number of detectable Galactic BBHs.

\end{abstract}


%5 \lmk   \frac{}{}\rmk 

\section{Introduction}
In September 2015, the Laser Interferometer Gravitational Wave Observatory (LIGO) discovered a gravitational wave source GW150914, identified as a merging binary black hole (BBH) (Abbott et al. 2016a).  The estimated individual masses in the source frame are relatively high $\sim 36M_\odot$ and $\sim 29M_\odot$. Including this discovery, Abbott et al. (2016c) conservatively estimated the comoving merger rate of BBHs in the range $R_V=2$-400 $\rm Gpc^{-3} yr^{-1}$.

The origin of this BBH is unclear, even though it is the first identified BBH.    Abbott et al (2016d) discussed the astrophysical implications of GW150914. The two heavy black holes are considered to be formed from massive stars whose metallicity is lower than 1/2 of the solar value. Abbott et al. (2016d) also mentioned the two main formation models for relatively massive BBHs; the isolated binary scenario ({\it e.g.} Dominik et 
al. 2012; Postnov 
\& Yungelson 2014; Kinugawa et 
al. 2014; Belczynski et 
al. 2015) and the dynamical scenario in  dense stellar environments ({\it e.g.} Kulkarni, Hut, 
\& McMillan 1993; Sigurdsson 
\& Hernquist 1993; Portegies Zwart 
\& McMillan 2000; Morscher et 
al. 2013; Tanikawa 2013; O'Leary, Meiron, 
\& Kocsis 2016). 

With the increase of the detectable volume and the observation time, more BBHs would be detected in the next $\sim5$ years. The redshift distribution of the detected BBH sample might be used to examine the two competing models, but would not be conclusive (Abbott et al. 2016d). The spin measurements of BBHs would be, in principle, very powerful to discriminate the two formation models, but would be challenging because of the limited measurement accuracy from GW signals   (Abbott et al. 2016d).

The initial  eccentricities of BBHs  would be largely different between the two models. For GW150914, Abbott et al. (2016b) quoted a preliminary constraint $e\lsim 0.1$ at 10Hz. But, due to the orbital circularization by gravitational radiation reaction, the residual eccentricity for both models would be too small to be measured with ground-based detectors. We will be able to constrain only  BBHs that are formed with  extremely high eccentricities, {\it e.g.} caused by a triple star interaction (Wen 2003;  Thompson 2011; Seto 2013; Antonini et al. 2014; Antognini et al. 2014; Samsing, MacLeod, 
\& Ramirez-Ruiz 2014).

Abbott et al. (2016d) briefly mentioned the detectability of a BBH in the local universe with eLISA that is sensitive to GWs in the 0.1-10mHz band (Amaro-Seoane et al. 2012). Considering the radiation reaction, the eccentricity of a BBH at $\sim 1$mHz can be at most $\sim 10^4$ times larger than the residual one at 10Hz. Therefore, eLISA might provide a crucial information to understand the formation processes of massive BBHs as GW150914. In this paper, we quantitatively discuss the prospects for the Galactic BBH search with eLISA, paying special attentions to its eccentricity measurement. 

For the fiducial model parameters of BBHs, we use the chirp mass $M_c=28M_\odot$, the total mass $M_t=65M_\odot$, and the comoving merger rate $R_V=150 \rm Gpc^{-3} yr^{-1}$, estimated from GW150914.

\section{Search for  Galactic BBH}

Since we are interested in the detection limit for a small eccentricity,  we hereafter assume $e\lsim 0.1$ for Galactic BBHs  in the eLISA band. 
This assumption significantly simplifies our formulation.

 An eccentric binary emits GWs at multiples of its orbital frequency $f_{\rm orb}$ (Peters 1964) as 
\beq
f_n\equiv f_{\rm orb}\times n~~~~(n=1,2,\cdots).
\eeq
For a  small eccentricity $e$, the $n=2$ mode is the  dominate component   with $A_2\propto e^0$ for its amplitude.  The $n=1$ and 3 modes are the sub-leading ones  with $A_1\propto e^1$ and   $A_3\propto e^1$. All other modes have negligible contributions with $A_n=O(e^2)$. 
Below, in stead of the orbital frequency $f_{\rm orb}$,  we mainly use the GW frequency $f_2=2f_{\rm orb}$  to characterize the orbit of a BBH.

Due to the GW emission, the orbit of a BBH shrinks, resulting in the GW frequency evolution  (Peters 1964)
\beqa
{\dot f_2}&=&\frac{96\pi^{8/3}G^{5/3}f_2^{11/3}M_c^{5/3}}{5c^5}\label{chirp}\\
&=&1.2 \times 10^{-16}\lmk \frac{f_2}{{\rm 0.5mHz}}\rmk^{11/3} \lmk \frac{M_c}{28M_\odot}\rmk^{5/3} {\rm sec^{-2}}.
\eeqa
From eq.(\ref{chirp}),
we can readily obtain the relation between the merger time $T_m$ and the GW frequency $f_2$ of a binary as 
\beq
T_m\propto f_2^{-8/3}M_c^{-5/3}.
\eeq
Similarly, the radiation reaction decreases the eccentricity of a binary  as a function of $f_2$ 
\beq
e\simeq e_i (f_2/f_{2i})^{-19/18},
\eeq
where $e_i\ll 1$ represents the initial eccentricity at $f_2=f_{2i}$ (Peters 1964).

Now, we consider the specific  BBH that has the highest GW frequency   in the Galaxy. Given the noise spectrum of eLISA at the low frequency regime, this BBH is expected to be the loudest one, namely  having  the largest  signal-to-noise ratio.  Below, we attach the subscript $L$ to represent quantities specific to the loudest Galactic BBH.

To begin with, we inversely apply the traditional conversion factor that relates the merger rate $R_G$ in a Milky Way Equivalent Galaxy  with the volume averaged merger rate $R_V$  (see {\it e.g.} Phinney 1991; Abadie et al. 2010) as 
\beq
R_G\simeq1.5\times 10^{-5}\lmk \frac{R_V}{150 {\rm Gpc^{-3} yr^{-1}}}\rmk {\rm yr^{-1}} .
\eeq
We should notice that this simple approximation might not be valid for some isolated formation scenarios with  short merger times (see Abbott et al. 2016d).

From eq.(4) and  the relation $T_{m,L} R_G\simeq1$ for the merger time of the loudest  Galactic BBH, its GW frequency is estimated as 
\beq
f_{2L}=0.5 \lmk  \frac{R_G}{\rm1.5\times 10^{-5} yr^{-1}}\rmk^{3/8} \lmk \frac{M_c}{28M_\odot}\rmk^{-5/8}{\rm mHz}.\label{f2}
\eeq
From the quadrupole formula (Peters 1964), the angular-averaged strain amplitude $A_2$ (for the $n=2$ mode) is  given as
\beqa
A_2&=&\frac{8G^{5/3} (\pi f_2)^{2/3}M_c^{5/3}}{5^{1/2}Dc^4}\\
& &=2.1\times10^{-20} \lmk \frac{D}{\rm 8kpc}\rmk^{-1} \lmk \frac{M_c}{28M_\odot}\rmk^{5/3} \nonumber\\
& & ~~~~~~\times \lmk \frac{f_2}{0.5{\rm mHz}}\rmk^{2/3}.\label{a2}
\eeqa

When we integrate this signal for an observational period $T_{obs}$, the accumulated signal-to-noise ratio $ SN_2$ becomes
\beqa
SN_2&=&\frac{A_2 T_{obs}^{1/2}}{h_n(f_2)}\\
&=&70 \lmk \frac{A_2}{2.1\times 10^{-20}} \rmk \lmk \frac{h_n(f_2)}{3\times 10^{-18}{\rm Hz^{-1/2}}} \rmk^{-1}\nonumber\\
& & ~~~~~~\times  \lmk  \frac{T_{obs}}{\rm 3yr}\rmk^{1/2}.\label{sn2}
\eeqa 
Here $h_n(f)$ represents the angular-averaged  strain noise of eLISA whose baseline value is  $3\times 10^{-18}{\rm Hz^{-1/2}}$ at 0.5mHz (Amaro-Seoane et 
al. 2012).  Around 0.1-1mHz,  the baseline eLISA will have the spectral shape $h_n(f)\psim f^{-\alpha}$ with   $\alpha\sim2$. The expression (11) shows that, for the fiducial model parameters,  the baseline eLISA is likely to detect the loudest Galactic BBH with a sufficient signal-to-noise ratio.

Actually, the scaling relations (\ref{a2}), (\ref{sn2}) (and similarly most of the expressions below) are valid for general binaries (only plugging-in  the reference parameters of the loudest BBH). But,  if we further use $f_{2L}\propto R_G^{3/8} M_c^{-5/8}$  derived specifically for the loudest Galactic BBH, we have the maximum signal-to-noise ratio as  $SN_{2L}\propto M_c^{(10-5\alpha)/8} R_G^{(2+3\alpha)/8}$. Interestingly, this expression is simplified as $SN_{2L}\propto M_c^0 R_G$ for $\alpha=2$, and does not depend on the fiducial  chirp mass $M_c$.  The mass dependence is canceled with the frequency dependence of the noise spectrum. We will show a similar scaling relation in \S 4.

So far, we have neglected astrophysical confusion noises. 
Applying the simple relation $\Omega_{GW}\propto f^{2/3}$ to the BBH GW background of  $\Omega_{GW}\simeq 3.8\times 10^{-9}$ at 25Hz (Abbott et al. 2016e), we obtain $\Omega_{GW}\simeq 4.4\times 10^{-12}$ at 1mHz. This is much smaller than the eLISA detector noise, corresponding to $\Omega_{GW}\gsim 10^{-9}$ below $\sim 1$mHz (Amaro-Seoane et 
al. 2012). The Galactic double white dwarf  (WD-WD) confusion noise is also expected to be sub-dominant in the frequency regime (Amaro-Seoane et 
al. 2012). 

Next, we discuss the parameter estimation for the detected  $n=2$ mode. For an observational period $T_{obs}\gsim 2$yr, eLISA can realize the following  resolutions (Takahashi \& Seto 2002, see also Cutler 1998)
\beq
\Delta {f}=3.1 \times 10^{-10} \lmk  \frac{SN}{70}\rmk^{-1}\lmk  \frac{T_{obs}}{3 \rm yr}\rmk^{-1},
\eeq
\beq
\Delta {\dot f}=6.1 \times 10^{-18} \lmk  \frac{SN}{70}\rmk^{-1}\lmk  \frac{T_{obs}}{3 \rm yr}\rmk^{-2}.
\eeq
In these expressions, we temporally  dropped the subscript $2$ for the mode, since these are valid  not only for the $n=2$ mode. Note that these results are obtained for simultaneous parameter fitting for a nearly monochromatic binary, including its frequency derivative $\dot f$ (Takahashi \& Seto 2002).

Therefore, the chip mass $M_c$ of the BBH can be estimated with the relative error of 
\beqa
\frac{\Delta M_c}{M_c}&\simeq& \frac{3 \Delta {\dot f_2}}{5 {\dot f_2}}\\
&\simeq& 0.03  \lmk  \frac{SN_2}{70}\rmk^{-1}\lmk  \frac{T_{obs}}{3 \rm yr}\rmk^{-2}\lmk \frac{M_c}{28M_\odot}\rmk^{-5/3}\nonumber\\
& & ~~~~~~\times  \lmk \frac{f_2}{0.5{\rm mHz}}\rmk^{-11/3}.
\eeqa
A long-term observation is crucial for the chirp mass estimation. Including the dependence $SN_2\propto T_{obs}^{1/2}$ in eq.(\ref{sn2}), we have the total time dependence $\Delta M_c \propto T_{obs}^{-5/2}$. For our fiducial model parameters, the chirp mass of the loudest Galactic BBH can be determined at $\sim 7\%$ accuracy, after a three-year observation.

The location of the BBH in the Galaxy might be also useful to study its formation model. 
At the low frequency regime $f_2\lsim 1$mHz, the direction of the BBH can be estimated by the annual amplitude modulation (Cutler 1998),  and the typical magnitude of the error box in the sky is given as  (Takahashi \& Seto 2002)
\beq
\Delta \Omega_s\sim 1.4\times 10^{-3} \lmk\frac{SN}{70}  \rmk^{-2}{\rm sr}.
\eeq
The estimation error for the distance $D\propto A_2^{-1} {\dot f_2}$ has a more complicated parameter dependence.

\section{Eccentricity measurement}

Now,  we discuss the possibility of the eccentricity measurement for the Galactic BBHs.  Our basic observational  strategy is to detect the sub-leading $n=1$ and 3 modes. Given the actual numerical coefficients of their amplitudes as well as the spectral shape of the detector noise, the $n=3$ mode is, by far, suitable for this  measurement (Seto 2001). It has the characteristic amplitude
\beq
A_3=\frac{9e}{4}A_2,
\eeq
and its angular-averaged signal-to-noise ratio $SN_3$ is given as
\beq
SN_3\simeq 7.5 \lmk \frac{e}{0.022}\rmk \lmk \frac{SN_2}{70}\rmk
\eeq
for the spectral index $\alpha=2$. Therefore, for the loudest BBH\footnote{Note that the higher post-Newtonian contribution to the $n=3$ mode would be sub-dominant  with the PN parameter $x=6\times 10^{-5}(M_t/64M_\odot)^{2/3}(f_2/{\rm 0.5mHz})^{2/3}$ for the Galactic BBH. We can expect an additional suppression for BBHs with similar masses as GW150914 (Arun et al. 2004).},  the eccentricity as small as $e\simeq 0.05$ could be detected by eLISA with the estimation error of 
\beq
\Delta e\sim \frac1{350}  \lmk \frac{SN_2}{70}\rmk^{-1}.
\eeq

Because of the orbital circularization  by the radiation reaction, the eccentricity $e\sim 0.02$ of a BBH  at $f_2=0.5$mHz will be decreased down to $e\sim 0.02(10/0.0005)^{-19/18}\sim5.8\times 10^{-7}$ at $f_2=10$Hz. For a BBH with the total mass $M_t\sim 65M_\odot$, this level can be hardly measured by ground-based detectors. Indeed, as mentioned earlier,  we only have a very rough bound $e\lsim 0.1$ for GW150914 (Abbott et al. 2016b).

Here we briefly comment on the potential misidentification  of the  $n=3$ mode with the GW emission lines from more abundant WD-WDs (Hils \& Bender 1990; Timpano, Rubbo 
\& Cornish 2006; Amaro-Seoane et 
al. 2012; Nissanke et 
al. 2012). 
The typical chirp mass of WD-WDs is $M_c\sim 0.3M_\odot$ (Farmer \& Phinney 2003) and only nearby WD-WDs could coincidently emit GW lines that might arise the confusion (see eq.(\ref{a2})). For this discrimination, parameters other than the frequency ({\it e.g.} sky direction) would be useful (Seto 2001).

If the $n=3$ mode is detected, we might additionally determine the total mass $M_t$ of the BBH (Seto 2001). Due to the PN effect, the pericenter of the binary precesses at the frequency 
\beq
\omega_p=\frac{1.0\times 10^{-7}}{1-e^2}\lmk \frac{M_{t}}{65M_\odot}\rmk^{2/3}\lmk \frac{f_2}{{\rm 0.5mHz}}\rmk^{5/3}\rm Hz.
\eeq
As a result, the frequencies of the $n=2$ and 3 modes satisfy
\beq
\frac32 f_2-f_3\simeq \frac{\omega_p}2.
\eeq

Therefore, by accurately measuring the frequencies $f_2$ and $f_3$, we can also estimate the total mass $M_t$ of the BBH with the error
\beqa
\frac{\Delta M_t}{M_t}&\simeq&3 \frac{\Delta f_3 }{\omega_p}\\
&\sim &0.09 \lmk\frac{SN_3}{7.5}  \rmk^{-1}\lmk \frac{f_2}{{\rm 0.5mHz}}\rmk^{-5/3}\lmk  \frac{T_{obs}}{3 \rm yr}\rmk^{-1}\nonumber\\
& & ~~~~~~\times \lmk \frac{M_{t}}{65M_\odot}\rmk^{-2/3}.
\eeqa

This argument was originally done for the GWs from Galactic double neutron stars (Seto 2001).
However,  unlike typical neutron stars, BHs could generally  have larger (non-dimensional) spin parameters and the orbital planes of BBHs could have larger precessions, due to the spin-orbit coupling (see {\it e.g.} Apostolatos et 
al. 1994; Gopakumar 
\& Sch{\"a}fer 2011). This might degrade the accuracy of the total mass estimation.

\if0
 But we will obtain more information about the spins of BBHs with the ground-based detectors (including the Einstein Telescope), before launch of eLISA. 
\fi

\section{Discussions}
So far, we have quantitatively discussed the signal analysis for the Galactic BBHs, based on the concrete  noise level of the baseline eLISA.
But, in reality, the sensitivity of eLISA would be a moving target and it would be interesting to consider how the Galactic BBH search changes with the detector noise level. To this end, simply neglecting the confusion background, we introduce the scaling parameter $X$ for the noise spectrum ($X=1$ for the baseline eLISA)
\beq
h_n(f)\propto X f^{-\alpha}.
\eeq
Using  $A_2\propto f_2^{2/3}M_c^{5/3}$ (fixing the distance $D$), we can derive a simple relation between the detection threshold $SN_{\rm 2th}$ and the corresponding GW frequency $f_{\rm 2th}$
\beq
SN_{\rm 2th}\propto A_2 T_{ obs}^{1/2}/h_n\propto f_{\rm 2th}^{2/3+\alpha}M_c^{5/3}X^{-1} T_{ obs}^{1/2}.
\eeq

Meanwhile, from eq.(4), the cumulative frequency distribution of the Galactic BBHs satisfies $N(>f_{2})\propto f_{2}^{-8/3}M_c^{-5/3} R_V$.  Then, for the spectral index $\alpha=2$ around the frequency  $f_{\rm 2th}$ and $D=8$kpc, we can derive a very simple relation for  the total number of detectable Galactic  BBHs 
\beqa 
 N(>f_{\rm 2th})&=&5.8 X^{-1} \lmk \frac{SN_{\rm 2th}}{12} \rmk^{-1} \lmk  \frac{R_V}{150 \rm Gpc^{-3} yr^{-1}} \rmk\nonumber \\
& & ~~ \times \lmk \frac{T_{ obs}}{3\rm yr} \rmk^{1/2}
\eeqa
without depending on $M_c$.
At the frequency regime 0.1-1mHz, the sensitivity of the baseline LISA is $\sim 10$ times better (namely $X\sim 0.1$) than that of eLISA (both with the spectral index $\alpha\simeq 2$). Therefore, LISA would detect $\sim 10$ times more Galactic BBHs above a given detection threshold. This could be a strong scientific motivation to pursue a better low-frequency sensitivity for a LISA-like missions.

\section{summary}

The advanced LIGO detectors discovered the gravitational wave source GW150914. This is the first confirmation of the existence of a  BBH. The masses of the two BHs are relatively high $\sim 30M_\odot$ and the formation process of this binary is unclear. 

We showed that, for our fiducial model parameters that are consistent with the current observation, eLISA is likely to detect at least one Galactic BBH around 0.5mHz with a sufficient signal-to-noise ratio. We found that eLISA also  has a  potential to measure its eccentricity as small as $e\sim 0.02$, corresponding to $e\sim 10^{-6}$ at 10Hz. With eLISA,  we might obtain a crucial information to understand the formation process of  relatively massive BBHs similar to GW150914.

This work was supported by JSPS (24540269, 15K05075) and
MEXT (24103006).

%%%%%%%%%%%%%%%%%%%%%%%%%%%%%%%%%%%%%%%%
%%%%%%%%%%%%%%%%%%%%%%%%%%%%%%%%%%%%%%%%

%\include{ref}

\end{document}